\documentclass[preprint,preprintnumbers]{revtex4}

\usepackage{graphicx}% Include figure files
\usepackage{dcolumn}% Align table columns on decimal point
\usepackage{bm}% bold math

\newcommand{\be}{\begin{equation}}
\newcommand{\ee}{\end{equation}}
\newcommand{\bea}{\begin{eqnarray}}
\newcommand{\eea}{\end{eqnarray}}

\begin{document}

% You should use BibTeX and revtex.bst for references
\bibliographystyle{apsrev}

\preprint{UAB-FT-528}

\title{On universality of the coupling of neutrinos to $Z$} 
% Force line breaks with \\

\author{Eduard Mass{\'o}} 
\email[]{masso@ifae.es}

%\homepage[]{Your web page}
%\thanks{}
%\altaffiliation{}
\affiliation{Grup de F{\'\i}sica Te{\`o}rica and Institut 
de F{\'\i}sica d'Altes
Energies\\Universitat Aut{\`o}noma de Barcelona\\ 
08193 Bellaterra, Barcelona, Spain}

%Collaboration name if desired (requires use of superscriptaddress
%option in \documentclass). \noaffiliation is required (may also be
%used with the \author command).
%\collaboration{}
%\noaffiliation

%\date{\today}% It is always \today, today,
             %  but any date may be explicitly specified

\begin{abstract}
We employ an effective Lagrangian approach and use LEP data
to place severe bounds on universality violations of the
couplings of $\nu_e$, $\nu_\mu$, and $\nu_\tau$ to the $Z$ boson. 
Our results justify the
assumption of universality in these couplings that is usually made,
as for example in the analysis of solar neutrinos
detected at SNO.
\end{abstract}

%\pacs{}% PACS, the Physics and Astronomy
                             % Classification Scheme.
%\keywords{Suggested keywords}%Use showkeys class option if keyword
                              %display desired
\maketitle

% body of paper here - Use proper section commands
% References should be done using the \cite, \ref, and \label commands

In the analysis of the observations on solar neutrinos at the 
Sudbury Neutrino
Observatory (SNO) \cite{sno} one makes the assumption that neutrinos 
interact with nucleons
and electrons according to the predictions of the electroweak 
Standard Model (SM). Of course, this assumption ought to be confirmed by
experiment. Although many of the basic neutrino properties predicted
by the SM and used in the SNO analysis have been tested in accelerator 
experiments, as we will argue this is not exactly true for all of them.

A crucial assumption in the SNO analysis 
involves the coupling of neutrinos to the neutral current. At SNO, 
neutral
current interactions induce the process 
$\nu_i + {\rm d} \rightarrow {\rm p}+{\rm n} + \nu_i$, where 
$i$ can be 
$e,\mu,\tau$, and $\nu_i + e^- \rightarrow \nu_i + e^-$ with 
$i=\mu,\tau$,
and they also participate in the elastic scattering
$\nu_e + e^- \rightarrow \nu_e + e^-$.
To deduce the actual fluxes of neutrinos reaching the Earth it is
assumed that $\nu_e$, $\nu_\mu$, and $\nu_\tau$ couple with
the same strenght to $Z$, i.e., that universality in the coupling
of neutrinos to the $Z$ boson holds. We will concentrate in the
observational evidence for this hypothesis.

Let us first make a very rough estimation of the precision level 
of SNO. 
Take for example the total flux measured with the neutral
current reaction at SNO \cite{sno},
\be
\phi_{\text{NC}}=5.09^{ + 0.63}_{ - 0.61} \,
\times 10^6~{\rm cm}^{-2} {\rm s}^{-1}
\label{phi}
\ee
where we have added their statistical and systematic errors in 
quadrature. The result has a relative error of about 12\%, and gives
us an idea of the SNO precision.
These are their very first results and, of course, with more 
statistics
and refinements, the relative error is going to decrease in the
near future.

To set the stage for our discussion, let us write the
Lagrangian that in the SM describes the interaction of matter with
the $Z$ boson,
\be
{\cal L}_{\text{NC}} = -\, \frac{g}{\cos\theta_w}\, 
J_\alpha^{\text{NC}} Z^\alpha
\label{LSM}
\ee
with $g$ the $SU(2)$ coupling and $\theta_w$ the weak mixing angle.
The neutral current $J^{\text{NC}}$ in the SM is given by
\be
J_\alpha^{\text{NC}} (\text{SM})= \frac{1}{2}\, \sum_i\,
[1+r_i]\, \bar \nu_i \gamma_\alpha \nu_i \ + \ \dots
\label{JNCSM}
\ee
where the dots refer to other particles and the sum is
over $i=e,\mu,\tau$. ($\nu_i$ is in fact $\nu_{Li}$; 
to simplify the notation we omit the left-handed $L$ subscript).
In (\ref{JNCSM}), $r_e$, $r_\mu$, $r_\tau$
are the usual radiative corrections arising in the SM.
At this radiative level, there are universality violations coming from
vertex corrections, but they are small. For instance, we have
\be
r_\tau - r_e \simeq \frac{\alpha}{4\pi} \log 
m_\tau^2/m_e^2 \simeq 0.009
\label{rad}
\ee
and even smaller for $r_\tau - r_\mu$ and $r_\mu-r_e$
(see for example Ref.\cite{radiative}). Certainly, this amount of
universality violation is not a concern for the SNO analysis. 
In our analysis we can also 
ignore it, since we will end up with upper bounds on universality 
violations that are larger that (\ref{rad}). 

In order to check the universality hypothesis in the neutrino
coupling to $Z$, we write the modified neutral current as
\be
J_\alpha^{\text{NC}} = \frac{1}{2}\, \sum_i\, [1+\Delta_i]\,
\bar \nu_i \gamma_\alpha \nu_i \ + \ \dots
\label{JNC}
\ee
The parameters
$\Delta_e$, $\Delta_\mu$, $\Delta_\tau$ are possible
deviations coming from physics beyond the SM. We will constrain 
these parameters using experiment.

Data from LEP constitutes a very precise test for the SM. 
The couplings of neutrinos to $Z$ are constrained
by the invisible $Z$ width, or equivalently,
in the determination of the number of neutrinos $N_\nu$.
A  combined fit to all LEP data gives
\be
N_\nu = 2.994 \pm 0.012
\label{Nnu}
\ee
that we take from the 2001 update of the PDG \cite{pdg}.
Each $\nu_i$ contributes 
\be
(1+\Delta_i)^2 \simeq 1 + 2\Delta_i
\label{change}
\ee
to $N_\nu$. Thus, the result (\ref{Nnu}) leads to the relation
\be
 |  \Delta_e+\Delta_\mu+\Delta_\tau |\, \leq\, 0.009
\label{Nnulimit}
\ee
To be conservative, we have taken the maximal deviation from 
$N_\nu=3$ in (\ref{Nnu}) to put the bound (\ref{Nnulimit}).
Also, here and in the following we work at first order in
$\Delta_i$.

If there is new physics that alter the $Z\nu\nu$ coupling
but respect universality, i.e., 
$\Delta_e=\Delta_\mu=\Delta_\tau$, then (\ref{Nnulimit})
implies that each individual $\Delta_i$ must be very small. However,
if universality is violated then me may have cancellations in
the sum (\ref{Nnulimit}) and, actually, there is no strict bound on
individual $\Delta_i$. Such possible 
cancellations cannot be banned from
first principles. The purpose of the present note is to
constrain the breaking of universality in $Z\nu\nu$ 
interactions.

The direct way to test universality in the neutrino coupling
to the $Z$ boson is through the analysis of the scattering 
$\nu_i + e^- \rightarrow \nu_i + e^-$. 
Available data on 
$\nu_\mu + e^- \rightarrow \nu_\mu + e^-$, together with
LEP results, allow to place the
limit \cite{vilain}
\be
 |\Delta_\mu| \leq 0.037
\label{numu}
\ee
However, the $\nu_e\nu_e Z$ coupling is known at a much worse level.
We have the experimental limit on universality violation 
\cite{dorenbosch}
\be
 0.13 \leq  \Delta_\mu-\Delta_e \leq 0.20
\label{nue}
\ee
and of course no limits involving $\Delta_\tau$. 
At the view of the SNO precision, this limit (and absence of
limit for the $\tau$ neutrino) is too loose to be useful.

We will now show that we can improve the limits on universality
working with effective Lagrangians.
The key point is the following. Deviations from the SM can
be treated by using effective Lagrangians. The general idea of
the effective Lagrangian approach is
that theories beyond the SM, emerging at some
characteristic energy scale $\Lambda$, have effects at
low energies $E \leq G_F^{-1/2}$, and these effects
can be taken into account by considering a Lagrangian
that extends the SM Lagrangian, ${\cal L}_{\rm SM}$:
\be
{\cal L} = {\cal L}_{\rm SM} + {\cal L}_{\rm eff} \ .
\label{L}
\ee
The effective Lagrangian ${\cal L}_{\rm eff}$ contains operators 
of
increasing dimension that are built with the SM fields including
the scalar sector, and is
organized as an expansion in powers of $(1/\Lambda)$.

The success of the electroweak SM at the level of quantum 
corrections can
be considered as a check of the gauge symmetry properties of the
model. To preserve the consistency of the low energy theory, with
a Lagrangian given by (\ref{L}), we will assume that
${\cal L}_{\rm eff}$ is $SU(2) \otimes U(1)$ gauge invariant.
Some of the problems that originate when dealing with non-gauge
invariant interactions have been discussed in \cite{rujula}.
The gauge-invariant operators that dominate at low
energies have dimension six and have been listed in 
 \cite{buchmuller}.

There are two classes of dimension-six operators that may 
originate
violations of universality in the neutrino sector of the neutral 
current:
\bea
{\cal A}_i &=& 
i\, [\Phi^\dagger\, {\cal D}_\alpha\, \Phi]\
[\bar L_i\, \gamma^\alpha\, L_i] \\
{\cal B}_i &=& 
i\, [\Phi^\dagger\, (\, {\cal D}_\alpha\, \vec \tau \, 
+ \,  \vec \tau\, {\cal D}_\alpha \, )\, \Phi] \cdot
[\bar L_i\, \gamma^\alpha \vec \tau\, L_i]
\label{AB}
\eea
Here $\Phi$ is the Higgs field and $L_i$ is the lepton isodoublet
\be
   L_i\, = \, 
   \pmatrix{l_i \cr \nu_i}_L
\label{Li}
\ee
where now $l_i$ are the charged leptons, and the subscript
$L$ is for left-handed.
The operators contain the covariant derivative,
\be
{\cal D}_\alpha = \partial_\alpha + 
i g\, \frac{\vec \tau}{2} \cdot \vec W_\alpha +
i g'\, \frac{Y}{2} B_\alpha
\label{D}
\ee
with the gauge bosons $\vec W,B$, the gauge couplings
$g,g'$, and the Pauli matrices $\vec \tau$ and the 
hypercharge $Y$.

The effective Lagrangian relevant for our purposes can now
be written as
\be
{\cal L}_{\rm eff} =
\sum_i\, \left( \frac{\alpha_i}{\Lambda_i^2}\, {\cal A}_i +
\frac{\beta_i}{{\Lambda'}_i^2}\, {\cal B}_i
\right)
\label{Leff}
\ee
where $\Lambda_i$, $\Lambda'_i$ are high-energy scales and
$\alpha_i$, $\beta_i$ are unknown strenght coefficients
accompanying the operators. Below the scale of spontaneous
symmetry breaking the effective Lagrangian (\ref{Leff})
induces contributions of the type shown in (\ref{JNC}).
Substituting 
\be
   \Phi \longrightarrow
\pmatrix{0\cr v/\sqrt{2}}
\label{Phi}
\ee
with $v^2 = 1/(\sqrt{2} G_F) \simeq (246\ {\rm GeV})^2$,
in (\ref{Leff}), we get
\be
\Delta_i = -a_i + b_i
\label{Delta}
\ee
corresponding to the contributions of the two operators,
\bea
a_i &=& \alpha_i\, \frac{v^2}{2\Lambda_i^2}\\
b_i &=& \beta_i\, \frac{v^2}{{\Lambda'}_i^2}
\label{ab}
\eea
It is clear that unless the combinations $-a_i+b_i$, for $i=e,\mu,\tau$
are equal, we will have universality violations in the
coupling of neutrinos to the Z boson. 

The operators in the effective Lagrangian (\ref{Leff})
have other effects at low energy. They contribute
to the couplings of the $Z$ boson to the charged leptons $l_i$.
Indeed we find,
\bea
J_\alpha^{\text{NC}} &=& \sum_i\, \left[ 
- \frac{1}{2} + \sin^2 \theta_w - \frac{a_i}{2} 
- \frac{b_i}{2} \right] \, \bar l_{iL}  \gamma_\alpha l_{iL} \\
&+& \sin^2 \theta_w\, \sum_i\, \bar l_{iR} \gamma_\alpha l_{iR}\, 
+ \dots
\label{JNCcharged}
\eea
where the dots indicate other particles than charged leptons.
We see that the couplings to right-handed charged leptons
$ l_{iR}$ are not modified.

Also, the charged current coupling to the charged $W$ gets
a contribution,
\bea
{\cal L}_{\text{CC}} &=& -\, \frac{g}{\sqrt{2}}\, 
J_\alpha^{\text{CC}} W^{+\alpha}\ +\ {\rm h.c.} \\
J_\alpha^{\text{CC}} &=& \sum_i\, [1+b_i]\,
\bar \nu_i \gamma_\alpha l_{iL} \, + \dots
\label{JJC}
\eea
where again the dots stand for the part involving other SM fields.
In the charged current sector there are also violations of
universality coming from radiative corrections in the SM framework,
but numerically they have at most the value shown in 
(\ref{rad}) and we will neglect them.

Our purpose is to constrain universality violations in the
couplings $\nu\nu Z$. We can reach our objective by considering
the constraints on the non-standard contributions to
the coupling of $Z$ to charged leptons (\ref{JNCcharged}) 
and on the
similar contributions in the charged current sector (\ref{JJC}).
The experimental information we need is taken from the 2001
PDG update that uses LEP data \cite{pdg}.

For example, in our scheme we have
\be
\frac{\Gamma(W\rightarrow \tau\nu)}{\Gamma(W\rightarrow e\nu)}\simeq
1 + 2(b_\tau - b_e)
\label{CCetautheory}
\ee
where we work at first order in $b_i$ and neglect lepton masses
in front of the $W$ mass. The experimental ratio
\be
\frac{\Gamma(W\rightarrow \tau\nu)}{\Gamma(W\rightarrow e\nu)}=
1.002 \pm 0.029
\label{CCetau}
\ee
leads to the bound
\be
2 |b_\tau - b_e| \leq 0.031
\label{CCetaulimit}
\ee
where again we conservatively use the maximal deviation from 1
in (\ref{CCetau}). We also consider
\be
\frac{\Gamma(Z\rightarrow \tau^+\tau^-)}{\Gamma(Z\rightarrow e^+e^-)}
\simeq
1 + \frac{1-2 s_w^2}{8s_w^4-4s_w^2+1 }\, 2 (a_\tau+b_\tau-a_e-b_e)
\label{NCetautheory}
\ee
($s_w=\sin \theta_w$) and use \cite{pdg}
 \be
\frac{\Gamma(Z\rightarrow \tau^+\tau^-)}{\Gamma(Z\rightarrow e^+e^-)}=
1.0020 \pm 0.0030
\label{NCetau}
\ee
which leads to
\be
 |a_\tau + b_\tau - a_e - b_e| \leq 0.0019
\label{NCetaulimit}
\ee

The inequality
\bea
|\Delta_\tau-\Delta_e| &=& |-a_\tau + b_\tau + a_e - b_e|\\
&\leq& |a_\tau + b_\tau - a_e - b_e| + 2 |b_\tau - b_e |
\eea
allows, from (\ref{CCetaulimit}) and (\ref{NCetaulimit}),
to get
\be
|\Delta_\tau-\Delta_e|\leq 0.033
\label{etaulimit}
\ee

This is our main result concernig the limit on universality
violation for $\nu_e$ and  $\nu_\tau$. We can do a totally
parallel exercise for $\nu_e$ and $\nu_\mu$, with the 
result
\be
|\Delta_\mu-\Delta_e|\leq 0.040
\label{emulimit}
\ee

We finally would like to mention that another assumption
of SNO is the absence of neutrino flavor changing neutral
currents (FCNC). Such exotic interaction would be a
contribution to $J^{NC}$ of the form
\be
\delta J^{NC}_\alpha = \Delta_{e\mu}\, 
\bar \nu_e \gamma_\alpha \nu_\mu \, + \, {\rm h.c.}
\label{FCNC}
\ee
and similar for $\nu_e\nu_\tau$ and $\nu_\mu\nu_\tau$.
The contributions of these FCNC parameters to $N_\nu$
have no interference, like the universality violations,
see (\ref{change}). It follows that now
cancellations are no longer possible. We can assume
non-zero $\Delta_{e\mu}$, $\Delta_{e\tau}$, and
$\Delta_{\mu\tau}$ and employ the experimental limit
on $N_\nu$ (\ref{Nnu}) to infer
\be
\left( (\Delta_{e\mu})^2 +  (\Delta_{e\tau})^2 +
(\Delta_{\mu\tau})^2 \right)^{1/2} \leq 0.095
\ee

In conclusion, we have placed 
severe constraints to universality violations of the
couplings of neutrinos to $Z$. 
We have introduced the parameters $\Delta_i$ in the neutral
current expression (\ref{JNC}) and showed which operators
in an effective Lagrangian approach may lead to universality
breaking. We have constrained the effects of this Lagrangian in
the sector involving charged leptons, and these constrains
have been used to reach  our numerical results 
(\ref{etaulimit}) and (\ref{emulimit}).
We have shown that universality
holds at the level of 4\% in the current that couples to $Z$.
Our results can be interesting in all the analysis making the
assumption of $\nu\nu Z$ universality, like in the solar neutrino 
experiment at SNO.

\begin{acknowledgments}
Work partially supported by the CICYT Research Project AEN99-0766, 
by the EU network on Supersymmetry and the Early Universe 
(HPRN-CT-2000-00152), and by the
{\it Departament d'Universitats, Recerca i Societat de la Informaci{\'o}},
Project 2001SGR00188.
\end{acknowledgments}

%\bibliography{apssamp}% Produces the bibliography via BibTeX.

\end{document}